\documentclass[twocolumn,showpacs,preprintnumbers,amsmath,amssymb,pre]{revtex4}

\usepackage{graphicx}
\usepackage{dcolumn}
\usepackage{bm}

\begin{document}
\title{Systematic characterization of thermodynamic and dynamical phase
behavior in systems with short-ranged attraction}
\author{P. Charbonneau\footnote{Current address: FOM
Institute for Atomic and Molecular Physics, Kruislaan 407, 1098
SJ Amsterdam, The Netherlands.}} \affiliation{Department of
Chemistry and Chemical Biology,Harvard University, 12 Oxford
Street, Cambridge, Massachusetts 02138, USA}
\author{D. R. Reichman}
\email{reichman@chem.columbia.edu} \affiliation{Department of
Chemistry, Columbia University, 3000 Broadway, New York, New
York 10027, USA}

\date{\today}
\begin{abstract}
In this paper we demonstrate the feasibility and utility of an
augmented version of the Gibbs ensemble Monte Carlo method for
computing the phase behavior of systems with strong, extremely
short-ranged attractions.  For generic potential shapes, this
approach allows for the investigation of narrower attractive
widths than those previously reported.  Direct comparison to
previous self-consistent Ornstein-Zernike approximation
calculations are made. A preliminary investigation of
out-of-equilibrium behavior is also performed. Our results
suggest that the recent observations of stable cluster phases
in systems without long-ranged repulsions are intimately
related to gas-crystal and metastable gas-liquid phase
separation.
\end{abstract}
\maketitle

\section{Introduction}
Colloidal systems with short-ranged interactions serve as a
minimal model of complex fluids used in a variety of
technological applications~\cite{russel:1991}.  While such
systems are important in the applied realm, their intrinsic
phase behavior is of great fundamental interest.  Upon adding
polymer to an otherwise uniform colloidal suspension, an
entropic depletion interaction is
induced~\cite{asakura:1954,vrij:1976}.  The range and strength
of this attractive interaction may be controlled by the length
and concentration of the added polymer.  Thus, exquisite
control may be experimentally exercised over such systems,
which allows for an exhaustive exploration of the
temperature-density phase diagram for nearly any range of
attractive
interaction~\cite{renth:2001,verma:1998,ramakrishnan:2002}.

At high volume fractions, depletion attractions have been shown to
induce an inverse melting behavior, such that the viscosity of a
hard-sphere suspension close to its colloidal glass transition is
significantly reduced~\cite{dawson:2002,pham:2002}. This suspension may be revitrified by
addition of still more polymer. Thus, two glassy phases appear to
exist at the same volume fraction: one induced by the repulsions and
one induced by strong attractions~\cite{dawson:2002}.  It has been speculated that the
attractive glass state at high volume fractions can be continuously
connected to a gel state at lower volume fractions~\cite{bergenholtz:2000}.  A major
difficulty with the connection between attractive glass and
colloidal gel is that phase separation often intervenes~\cite{cates:2004,foffi:2005}. Phase
separation into colloid rich and colloid poor regions may become
anomalously slow if the density of the colloid rich region is close
to the density of the uniform attractive glass at high volume
fractions.  This process leads to weak colloidal gels whose
connection with near-equilibrium high volume fraction attractive
glasses is complicated by the interplay of vitrification and phase
separation~\cite{verhaegh:1997,manley:2005}. Thus, the added dimension of out-of-equilibrium behavior
may greatly increase the complexity of the various phases that may
be observed in such systems.

In addition to colloidal gels induced by phase separation, various
cluster phases have been recently experimentally
observed~\cite{segre:2001,lu:2006,sedgwick:2005}.  If charge resides
on the colloids, then equilibrium microphase clusters may form due
to the competition between short-ranged attraction and long-ranged
charge repulsion~\cite{sciortino:2004,sciortino:2005,deCandia:2006}.
Interestingly, recent experiments that utilize sufficient salt to
screen the charge on the colloids still show the existence of
relatively stable large clusters.  The existence of such clusters
even in the absence of repulsion has been suggested
theoretically~\cite{kroy:2004}. Lu {\em et al}.~\cite{lu:2006} have
observed large, relatively compact clusters at low volume fraction
and low ($\sim 1-2 k_{B}T$) attraction strength.  Sedgwick {\em et
al}.~\cite{sedgwick:2005} have observed a ``bead phase'' of clusters
that appears to exist only along a portion of the metastable
gas-liquid binodal.

Clearly, the investigation of the nature of gel and cluster
phases via computer simulation is complicated by the need for a
precise characterization of the equilibrium phase diagram.
Qualitatively, the role of short-ranged attractions in widening
the gas-solid coexistence gap is well
known~\cite{gast:1983,hagen:1994}.  The characterization of
various phase boundaries for general short-ranged potentials,
however, has been performed by integral equation
methods~\cite{foffi:2002} by a mix of Monte Carlo and analytic
expansions~\cite{asherie:1996,lomakin:1996}, on systems with a
moderately
short attraction range~\cite{tenwolde:1997,bolhuis:2002,pagan:2005,liu:2005,vliegenthart:1999,%
shukla:2000,chang:2004,pellicane:2003,tavares:1997,dijkstra:1999,fortini:2005},
or short-ranged attractive systems with somewhat artificial
forms that do not lend themselves to dynamical
studies~\cite{miller:2003,miller:2004}. In this work, we will
combine several existing computational methodologies to produce
a direct Monte Carlo approach that is capable of yielding
essentially exact phase behavior over a wide range of the
thermodynamic parameter space, for systems with extremely
short-ranged attractive interactions.

In particular, we demonstrate the general applicability of a
form of the Gibbs ensemble Monte Carlo
(GEMC)~\cite{panagiotopoulos:1992} approach augmented with
improved sampling techniques for the calculation of phase
behavior in systems that mimic colloidal suspensions with
short-ranged attractions. In addition, our approach allows for
the direct sampling of configurations without additional
\emph{a priori} knowledge when the phase behavior is complex,
such as in systems that exhibit microphase
separation~\cite{charbonneau:2006}. We use the results gleaned
from this implementation of GEMC to compare with several
techniques, including the approximate but powerful
self-consistent Ornstein-Zernike approximation (SCOZA)
approach~\cite{foffi:2002} and exact approaches such as
thermodynamic integration~\cite{dijkstra:2002}.  With these
results in hand, we make a preliminary study of the
nonequilibrium phase behavior of systems presumably similar to
those studied by Lu {\em et al}.~\cite{lu:2006} and Sedgwick
{\em et al}.~\cite{sedgwick:2005} in the regimes where cluster
phases might be expected.  Our paper is organized as follows.
In Sec.~\ref{sect:simu}, we discuss the application of our GEMC
methodology to systems with short-ranged attractions, and we
outline the systems that are studied in this work. In
Sec.~\ref{sect:scoza}, we compare the results of our GEMC to
previously published results. In Sec.~\ref{sect:dyna} we study
various aspects of the out-of-equilibrium behavior of systems
similar to those published in recent experiments. Our studies
are facilitated by the precise knowledge of the phase diagram
afforded by the Monte Carlo method developed in this work. In
Sec.~\ref{sect:conclu} we conclude.

\section{Simulation techniques}
\label{sect:simu}
\begin{figure}[ht]
\center{\includegraphics[width=0.9\columnwidth]{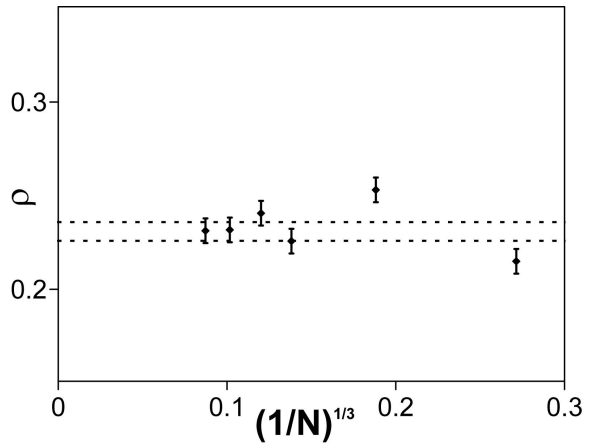}}\\
\caption[GEMC gas-crystal coexistence finite size study]{GEMC $b=30$
gas density for the gas-crystal coexistence at $T=0.33$ for
different crystal slab dimensions $N$. Within error bars, the bulk
limit (within the GEMC error interval) is attained when the
crystal has $\gtrsim500$ particles.} \label{fig:finsize}
\end{figure}

\begin{figure*}[ht]
\hspace{0.5in}\textbf{(a)\hfill{(b)}}\hspace{3in}\vspace{-0.2in}\\
\center{\hspace{0.5in}\includegraphics[scale=.25]{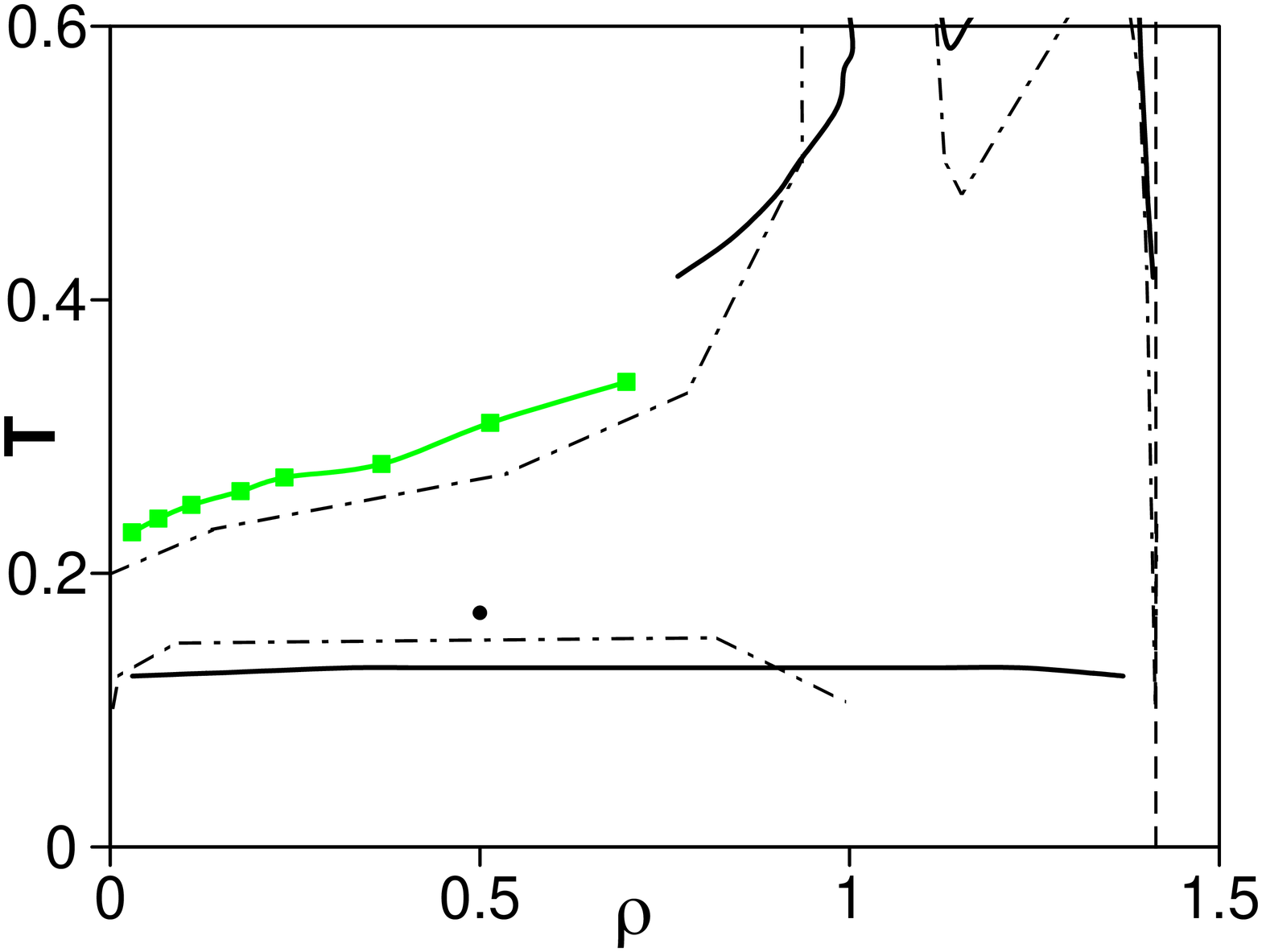}\hfill%
\includegraphics[scale=.25]{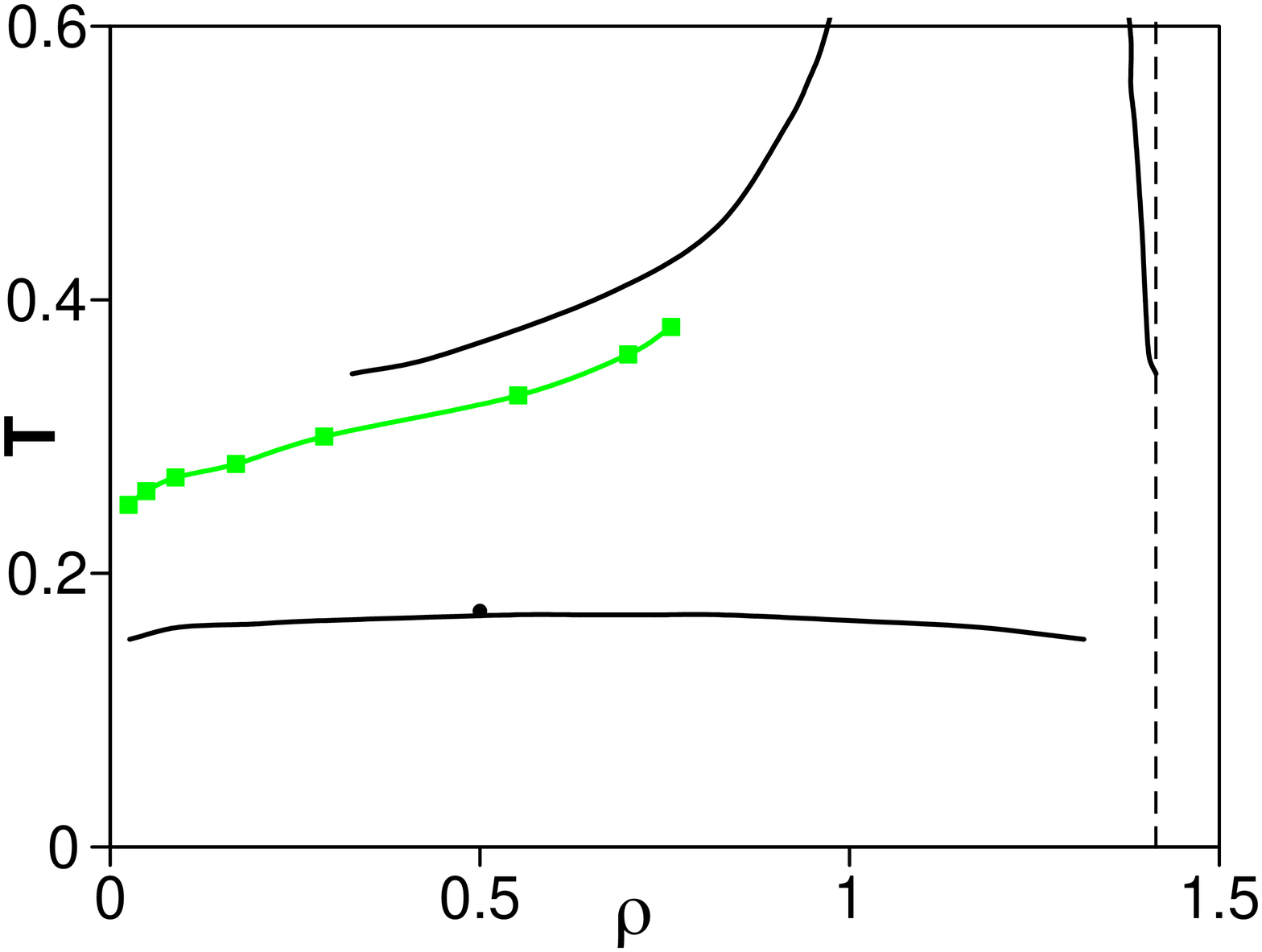}\hspace{0.5in}}\vspace{-0.2in}\\
\hspace{0.5in}\textbf{(c)\hfill{(d)}}\hspace{3in}\vspace{-0.2in}\\
\center{\hspace{0.5in}\includegraphics[scale=.25]{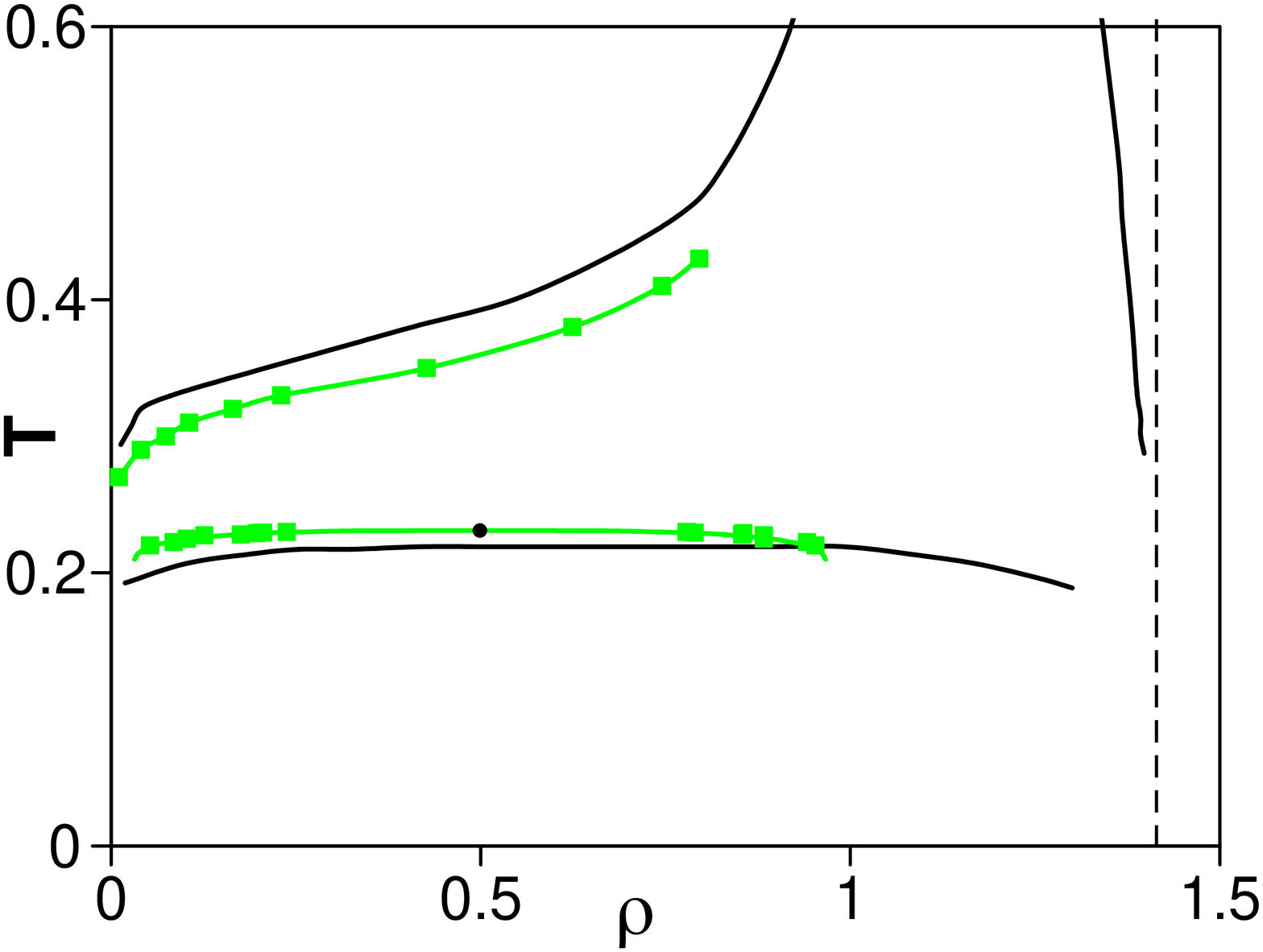}\hfill%
\includegraphics[scale=.25]{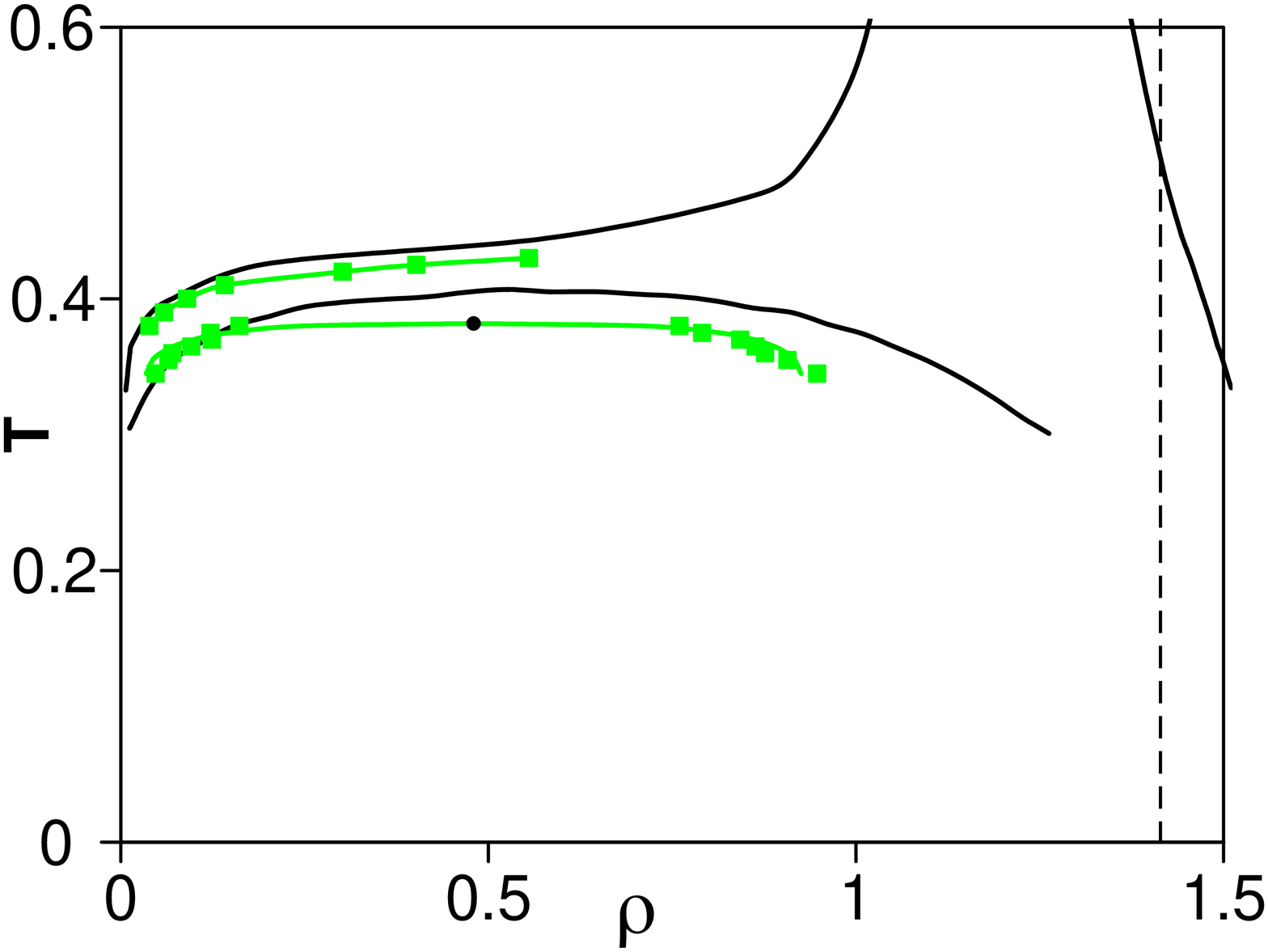}\hspace{0.5in}}\vspace{-0.2in}\\
\caption[Comparison of SCOZA with GEMC phase diagrams for
short-ranged hard-core Yukawa potenials]{(Color online)
Comparison of GEMC results (green squares) with
SCOZA~\cite{foffi:2002} (solid black lines) phase diagram
calculations for the hard sphere Yukawa potential with $b=100$,
$60$, $30$, and $7.5$ for panel (\textbf{a}), (\textbf{b}),
(\textbf{c}), and (\textbf{d}), respectively. The gas-crystal
binodal for a given density is found at higher temperatures
than the gas-liquid binodal in all systems studied. The dashed
line indicates the crystal close-packing limit. For $b=100$,
results from Ref.~\cite{dijkstra:2002} (dash-dotted line) are
also included. For $b=100$ and $60$, only the GEMC gas-crystal
binodal is shown. The black dots are the critical point
predicted by extended corresponding state, which are $T_c=
0.1711$, $0.1725$, $0.231$, and $0.382$, respectively. The line
between the squares is a guide for the eye in the gas-crystal
case; refer to the text for the gas-liquid case. The error bars
due to density fluctuations are smaller than the symbol sizes.}
\label{fig:yukph}
\end{figure*}

Various potential shapes have been suggested to capture the
phenomenology of short-ranged attractive systems. Two commonly used
forms are the generalized Lennard-Jones (LJ) potential
\begin{equation}
U(r)=4\epsilon
\left[\left(\frac{\sigma}{r}\right)^{2n}-\left(\frac{\sigma}{r}\right)^n\right]
\end{equation}
with $n>12$~\cite{hasegawa:1997} and the hard core plus attractive
Yukawa potential
\begin{eqnarray}
U(r)=
\left\{\begin{array}{cc}
  \infty & r<\sigma \\
  -\epsilon\frac{e^{-b\frac{(r-\sigma)}{\sigma}}}{r/\sigma} & r\geq\sigma
\end{array}\right.
\end{eqnarray}
with $b\gtrsim6.05$~\cite{foffi:2002}. Both are used in the
current work. Results are reported in reduced units.
Temperatures are in units of the well depth $\epsilon$,
distances are in units of the particle diameter $\sigma$, and
time is rescaled by $(\epsilon/m \sigma^2)^{1/2}$ where $m$ is
the mass of the particles.

GEMC with single particle displacement and exchange as well as
volume exchange is used for the metastable gas-liquid
equilibrium for two boxes of 256 particles (500 for $b=60$ and
LJ potentials). Finite size effects were checked by comparing
the results to that of a square-well potential with a width of
$0.25\sigma$~\cite{pagan:2005}. As the work was being
completed, a paper with similar methodology (applied to square
well fluids) appeared, and thus more details can be found
there~\cite{liu:2005}. In the current work, $50\%$ of the moves
are particle swaps, $49.5\%$ are particle displacements, and
the rest are volume interchanges, while equilibrium and
production runs involve at least $10^6$, and sometimes reaching
$10^7$ Monte Carlo cycles. When solid nucleated in the liquid
box, only the intermediate results are retained for analysis.

For the gas-crystal equilibrium, the methodology developed by
Chen and Siepmann is followed~\cite{chen:2000,chen:2001}. This
greatly increases the efficiency of GEMC in this regime. The
boxes are configured such that one box contains only vapor and
the other a slab of solid surrounded by vapor. In addition to
basic GEMC moves ($47\%$ particle displacements, $40\%$
particle swaps, $3\%$ symmetric and asymmetric volume
exchanges) aggregation-volume bias~\cite{chen:2000} and its
generalization to exchanges between two boxes ($5\%$ each) are
performed. For these simulations initial system sizes are 256
particles for the gas phase, 864 for the solid slab, and again
a minimum of $10^6$ Monte Carlo cycles are performed for
equilibration and production.

All dynamical results are performed with the Lennard-Jones
potential using standard molecular dynamics integration for
systems of size $N=5324$ and with an integration step $\Delta
t=0.001$. Cooling is done by velocity resampling every $10^5$
steps starting from a liquid configuration. We do not expect a
strong influence of the type of dynamics on the qualitative
results~\cite{foffi:2005b}, so further studies with other
dynamical protocols are not performed.

\subsection{Finite-size effects}
\label{sect:finsize} An obvious source or error exists in the
approach we have employed to determine the gas-crystal
coexistence line. In particular, the standard use of GEMC for
gas-crystal coexistence would utilize a box of pure solid and a
box of pure vapor with particle exchanges between the two
boxes. In such a configuration, exchanges between boxes would
be prohibitively infrequent.  We have instead used the Chen and
Siepmann approach where the solid box is replaced by a box
containing a solid slab surrounded by vapor~\cite{chen:2001}.
From the microscopic perspective, several possible finite-size
effects arise. First, the crystal block might show anisotropy
in its vapor pressure depending on the selection of the grain
surface exposed to the gas. However, this is expected to be
very small for spherically symmetric potential shapes. Second,
at high densities, crystal surface fluctuations or other image
effects may become significant. This was avoided by working at
relatively low gas densities and by using a box of sufficient
dimensions to eliminate periodic artifacts. From the
thermodynamic standpoint, the Chen-Siepmann procedure can lead
to biased results due to the fact that the free energy of the
solid box now has a contribution from the surface. In
principle, this error should scale as $\sim 2\sigma_s/L$ where
$\sigma_s$ is the surface tension and $L$ the crystal size. In
previous studies, Chen and Siepmann have shown that their
results are in quantitative agreement with the known
gas-crystal coexistence behavior of several simple
systems~\cite{chen:2001}. While they did not perform a study of
finite size effects, their results suggest that, in at least
some instances, it is indeed possible to obtain quantitative
results efficiently with a modified GEMC approach that allows
for metastable equilibrium via ``coexistence'' between a gas on
the one hand and a gas-crystal mixture on the other.  However,
the convergence as a function of system size should be
performed at each temperature and for each system, in order to
ensure accurate results within the modified GEMC approach of
Chen and Siepmann. Here, we carry out such a study.

In Fig.~\ref{fig:finsize} we show the gas density obtained for
$b=30$ and $T=0.33$ for a variety of different system sizes.
After some noticeable damped oscillatory behavior for small
system sizes, the value remains within the GEMC error width. We
consider the density to be converged when the result has
saturated within a band of densities whose width is that of the
error bars of the GEMC procedure itself for a given size of the
solid slab. The nontrivial convergence behavior hints at a
possible lattice periodicity effect (e.g., the period of the
behavior might be related to the number of particles in a layer
of solid), but a more detailed analysis is left for a later
study.  The results presented in Fig.~\ref{fig:finsize} are
representative of the results found for other values of $b$ and
$T$ that are presented in this work.

Thus, we are confident of the accuracy of the results for the
gas-crystal lines presented in Figs.~\ref{fig:yukph} and
\ref{fig:ljph}. While it is true that the method used here for
gas-crystal coexistence can induce systematic error, it appears
that such errors can be kept to a controllable size without
undue computational effort, at least in the systems we have
investigated. It should also be noted that while approaches
such as thermodynamic integration do not suffer from the
potential systematic bias discussed in this subsection, they
still may lead to inaccurate results if a good reference system
is not used (as could the case for extremely short-ranged
attractive potentials) or if the systems utilized are too
small. This will be discussed further in the next section.

\section{Comparison with SCOZA}
\label{sect:scoza}

\begin{figure}[ht]
\center{\includegraphics[scale=.25]{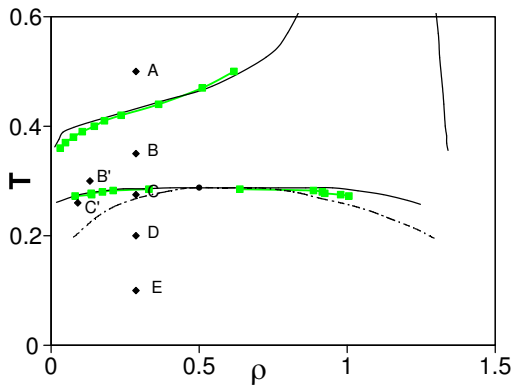}}\\
\caption[GEMC Lennard-Jones phase diagram for $n=50$]{(Color online) GEMC phase
diagram (green squares) for the LJ potential with $n=50$ is shown
superimposed with shifted SCOZA (solid black lines) results for
$b=30$. The lines between the squares are a guide for the eye. The SCOZA spinodal line is indicated by a dash-dotted line.
See text for details on the fit. The error bars from density
fluctuations are smaller than the symbol sizes.} \label{fig:ljph}
\end{figure}

\begin{figure}
\center{\hspace{-0.1in}\includegraphics[scale=.3]{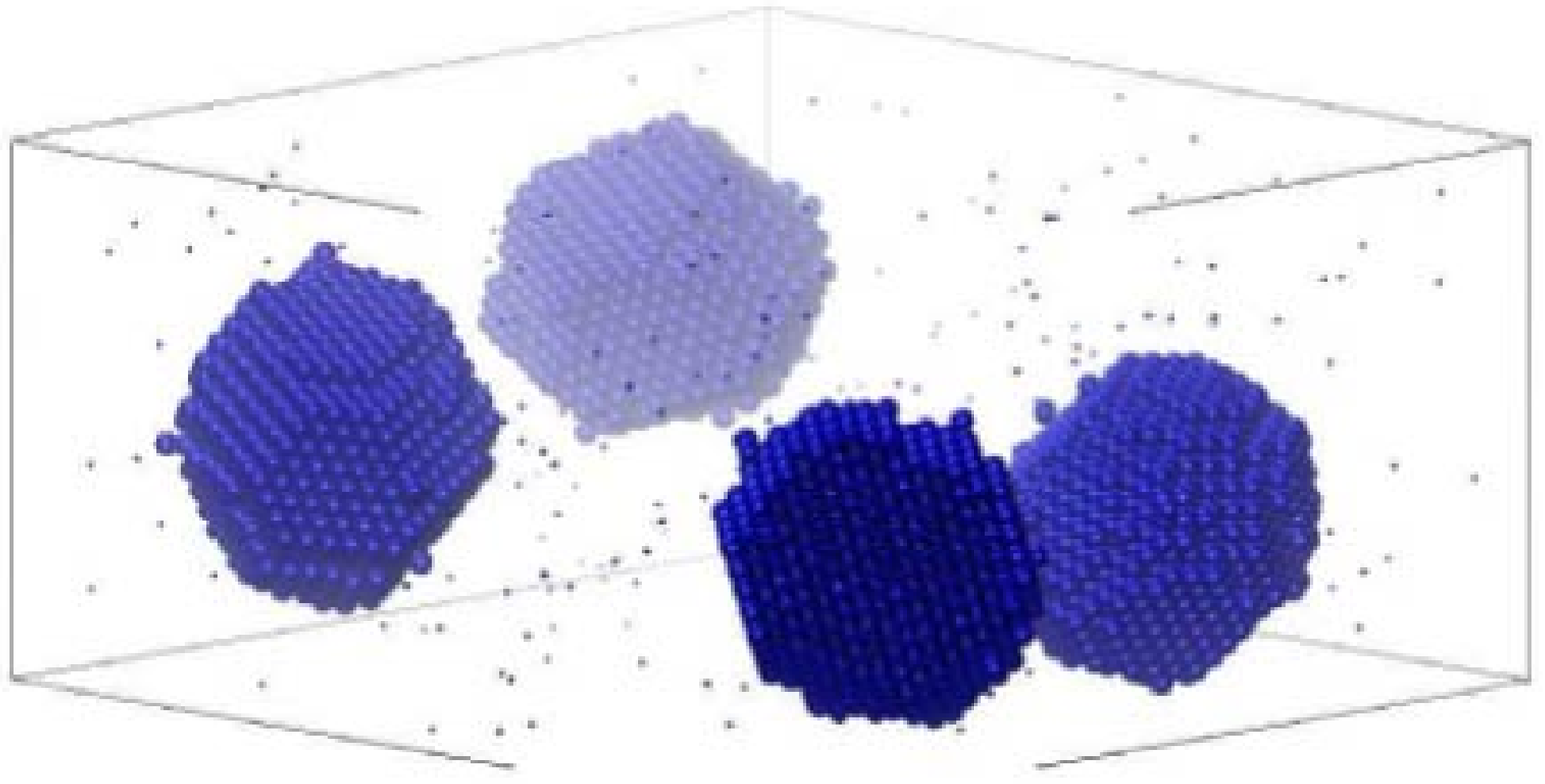}\\\includegraphics[scale=.33]{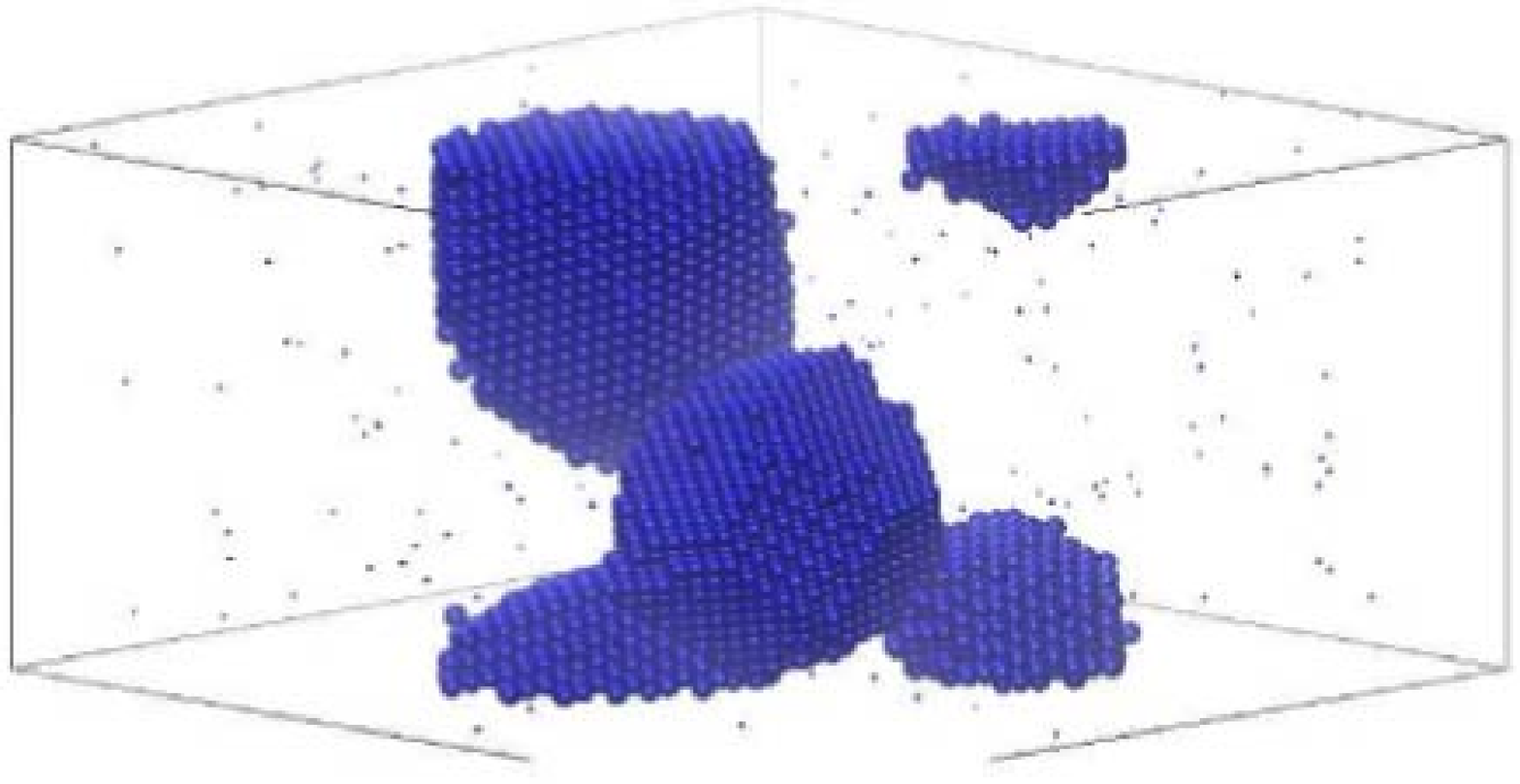}
} \caption[Aggregation of seeded nuclei in the gas-crystal
coexistence region]{(Color online) The system at point B' in
Fig.~\ref{fig:ljph} once nucleated at (a) early times and (b)
later times.
The details of the artificial seeding process are discussed in the text.
} \label{fig:ljclus}
\end{figure}

Various aspects of the thermodynamic properties of systems with
short-ranged attractions have been studied in the last decade~\cite{hagen:1994,%
tenwolde:1997,vliegenthart:1999,%
pagan:2005,liu:2005,foffi:2002,shukla:2000,miller:2004}. One of
the interesting features of such systems is the possibility of
metastable gas-liquid coexistence below the gas-crystal
equilibrium region. However, no verification of integral
equation predictions, other than in the Baxter limit of an
infinitely narrow potential, has been made in extremely
short-ranged attractive systems~\cite{miller:2004}. SCOZA is
generally viewed as being among the most accurate integral
equation methods available~\cite{hoye:1977,hoye:1984}.
Predictions of phase diagrams for short-ranged potentials with
hard-core Yukawa have been made using SCOZA and SCOZA-based
hybrid methods~\cite{foffi:2002}. However, these results have
only been compared with simulation results slightly over the
limit of metastability with
$b=7$~\cite{hagen:1994,shukla:2000,foffi:2002} for the
gas-liquid binodal and up to $b=9$ for the gas-solid
binodal~\cite{hagen:1994}. The Baxter limit of infinitely small
interaction range has also been reported via
simulation~\cite{miller:2004}, but only for the gas-liquid
binodal. For colloidal depletion interactions, ranges of the
order of $0.25\sigma$ to $0.01\sigma$ are of more interest,
corresponding to $b=20-100$ for the hard-core attractive
Yukawa. Simulated phase diagrams for systems with an
interaction range as low as $0.15\sigma$ for square
well~\cite{pagan:2005,liu:2005}, and $\gtrsim0.08\sigma$ for
Derjaguin-Landau-Verwey-Overbeek (DLVO)~\cite{pellicane:2003},
LJ-based~\cite{vliegenthart:1999,chang:2004,tenwolde:1997},
Asakura-Oosawa pair
potential~\cite{tavares:1997,dijkstra:1999,fortini:2005} have
been reported, but none of these can easily be quantitatively
compared with the SCOZA predictions, other than the cases
presented in Ref.~\cite{dijkstra:2002}.

Calculated phase diagrams for the hard core attractive Yukawa
potential are shown in Fig.~\ref{fig:yukph}. The metastable
gas-liquid density-temperature binodal was fitted using
$|\rho-\rho_c|=A|T-T_c|-B|T-T_c|^\beta$, where
$\beta=0.3258$~\cite{vega:1992} and the critical temperature
$T_c$ used in the fitting was extracted from the result of the
Baxter limit~\cite{miller:2003} and mapped to this system by
equating second virial coefficients as done for similar
systems~\cite{foffi:2005b}. Due to the flatness of the curve,
critical densities $\rho_c$ are harder to pinpoint. A value of
$\rho_c\sim0.50(5)$ appears to fit all of our simulated
systems. No unique set of $A$ and $B$ parameters fit the
binodal through the entire temperature range of our
simulations, but a rather reasonable agreement was obtained
nonetheless for fixed $A$ and $B$. The results for the
hard-core Yukawa potential for $b=7$ and $b=25$ for $T_c$ and
$\rho_c$ are consistent with those of Ref.~\cite{dijkstra:2002}
for nearby values of $b$.

For the gas-solid binodal, a simple spline fit was made to the
gas branch. Densities on the solid side are indistinguishable
from that of close packing within measured precision. Much
larger system sizes would be required to obtain a sufficiently
precise measurement. This leads us to suspect the solid phase
density results obtained by a recent paper are due to an
erroneous density calculation scheme for these
systems~\cite{liu:2005}. However, this is not the part of the
phase diagram that is of interest here, and we will thus
satisfy ourselves with a straight line at the close-packing
density. Finally, the range of densities available for
simulations did not allow for the study of solid-solid
coexistence, known to take place in systems with an extremely
short range of interaction~\cite{bolhuis:1994a,bolhuis:1994b}.

Comparing the phase diagrams of two different potential shapes
with a similar range of attraction ($\sim 0.1\sigma$), namely
the LJ with $n=50$ and the hard core attractive Yukawa
potential with $b=30$, we observe that the interaction range is
a good measure of gas-liquid phase separation metastability:
the gap between the metastable gas-liquid critical point and
the gas-solid binodal is $T\approx0.15$ in both cases. After
shifting the phase diagram so as to match critical points,
SCOZA predictions are seen to match the LJ simulation results
just as well in Fig.~\ref{fig:ljph}. This is in the spirit of
the extended corresponding state principle discussed above.

Comparison of GEMC simulation with SCOZA shows rather good agreement
for the gas-solid binodal, especially for the wider interaction
ranges of $b=7.5$ and $b=30$. For the narrower ranges of $b=60$ and
$100$, we reach densities beyond the range numerically attainable
for SCOZA calculations~\cite{foffi:2002}. Also, though the curves have
similar shapes, the simulation results are shifted to lower
temperatures with respect to the SCOZA predictions. On the high
density side, phase boundaries would only be accessible through the GEMC
methodology using much larger systems due to the fact that small
blocks of crystal tend to melt completely at high vapor pressures.
The pathological SCOZA predictions of solids beyond the
close-packing density indicates the limitation of such an approach
in high density regimes.

GEMC metastable gas-liquid binodals for $b=7.5$ and $30$ also
agree well with SCOZA predictions, especially for the low
density branch. It is difficult to reach lower temperatures as
solid nucleation becomes facile in metastable finite-sized
systems. This limits the range of temperatures where the
metastable binodal data can be obtained. Furthermore, this
approach breaks down for narrower interaction ranges when the
flatness of the binodal does not allow for a sufficient
separation of Monte Carlo time scales before the crystal
nucleates. Obtaining spinodal curves by simulation would
require precise dynamical measurement upon cooling and as its
meaning is loosely defined in any case~\cite{foffi:2002}, no
direct calculation of the spinodal was attempted. For further
reference the SCOZA spinodal was included in the LJ phase
diagram in Fig.~\ref{fig:ljph}~\cite{foffi:2002}.

Finally, we conclude this section with a brief discussion of
the comparison of our results for a very narrow attraction
range ($b=100$) with that of thermodynamic integration
techniques~\cite{dijkstra:2002}. In principle, thermodynamic
integration is the preferred method, since it does not have a
systematic bias such as, for example, that discussed in
Sec.~\ref{sect:finsize}. However, in some important cases the
methods presented here have some advantages over this direct
approach. In particular thermodynamic integration may be
difficult to perform if the repulsive portion of the potential
is not of the hardcore variety or if a suitable reference
system cannot be found.  More importantly, in cases where the
morphological characteristics of the phases are not known (such
as in complex microphases or domain-forming
systems~\cite{charbonneau:2006}) it is of great advantage to
have a direct Monte Carlo approach that does not need to make
use of presupposed information. In Fig.~\ref{fig:yukph}a we
compare the thermodynamic integration results of
Ref.~\cite{dijkstra:2002} to that of our GEMC approach.  While
the results of Ref.~\cite{dijkstra:2002} are consistent with
our results for smaller values of $b$, here, a difference of
about $10\%$ can be seen in the gas-crystal coexistence
temperatures at low densities.  We believe, based on an
analysis of the type given in Sec.~\ref{sect:finsize} that our
results should be quantitatively accurate. Interestingly, a
similar discrepancy of the same magnitude and direction exists
between the location of the gas-liquid binodal line and the
location of the critical point as found by the extended
corresponding state predictions. It is possible that the small
system size ($N=108$) used in the thermodynamic integration
study is one source of these discrepancies. It would be of some
interest to examine these issues further in a future study.


\section{Nonequilibrium kinetics in various regions of the phase
diagram} \label{sect:dyna}
\begin{figure*}
\center{\includegraphics[scale=.29]{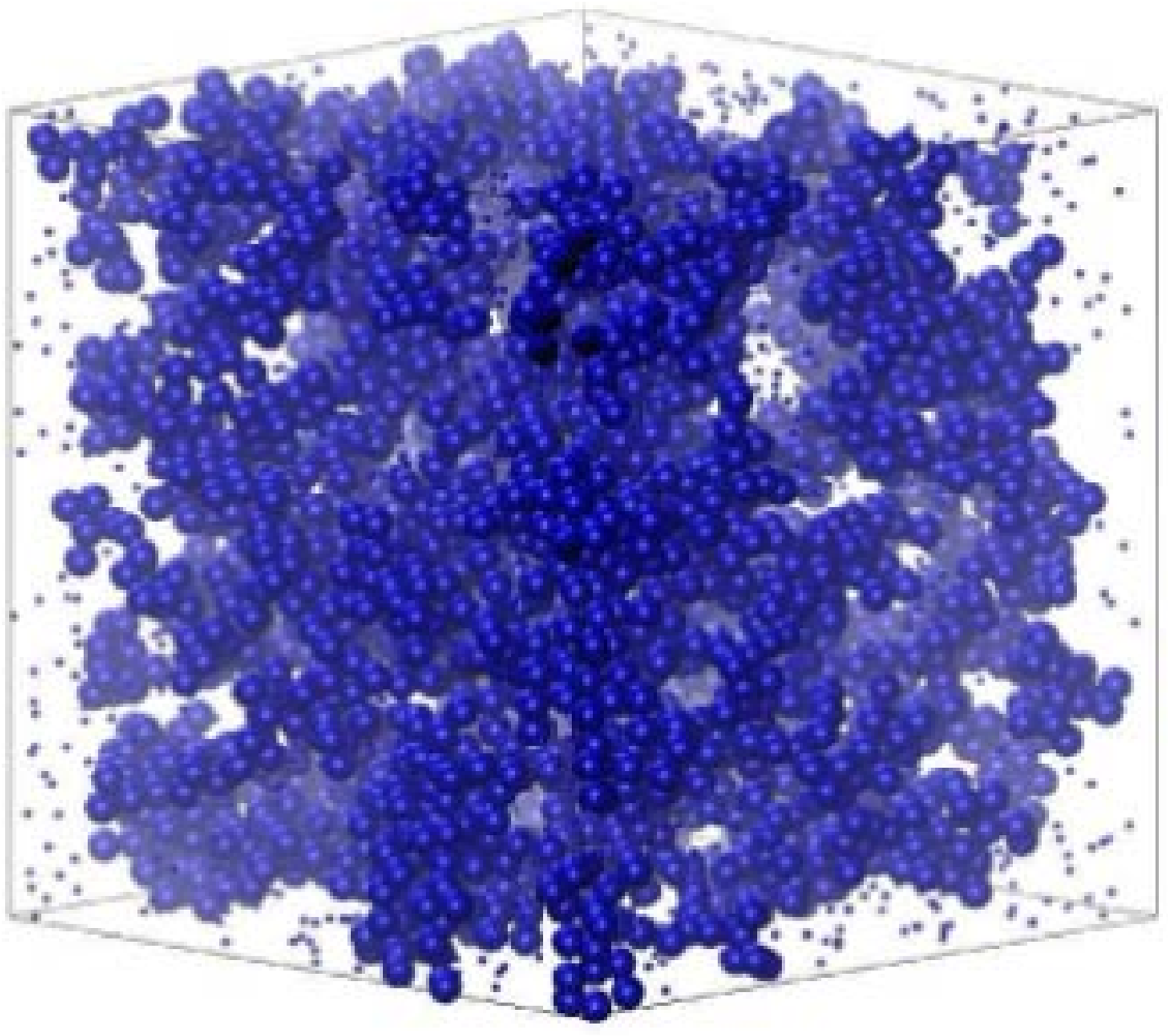}\includegraphics[scale=.28]{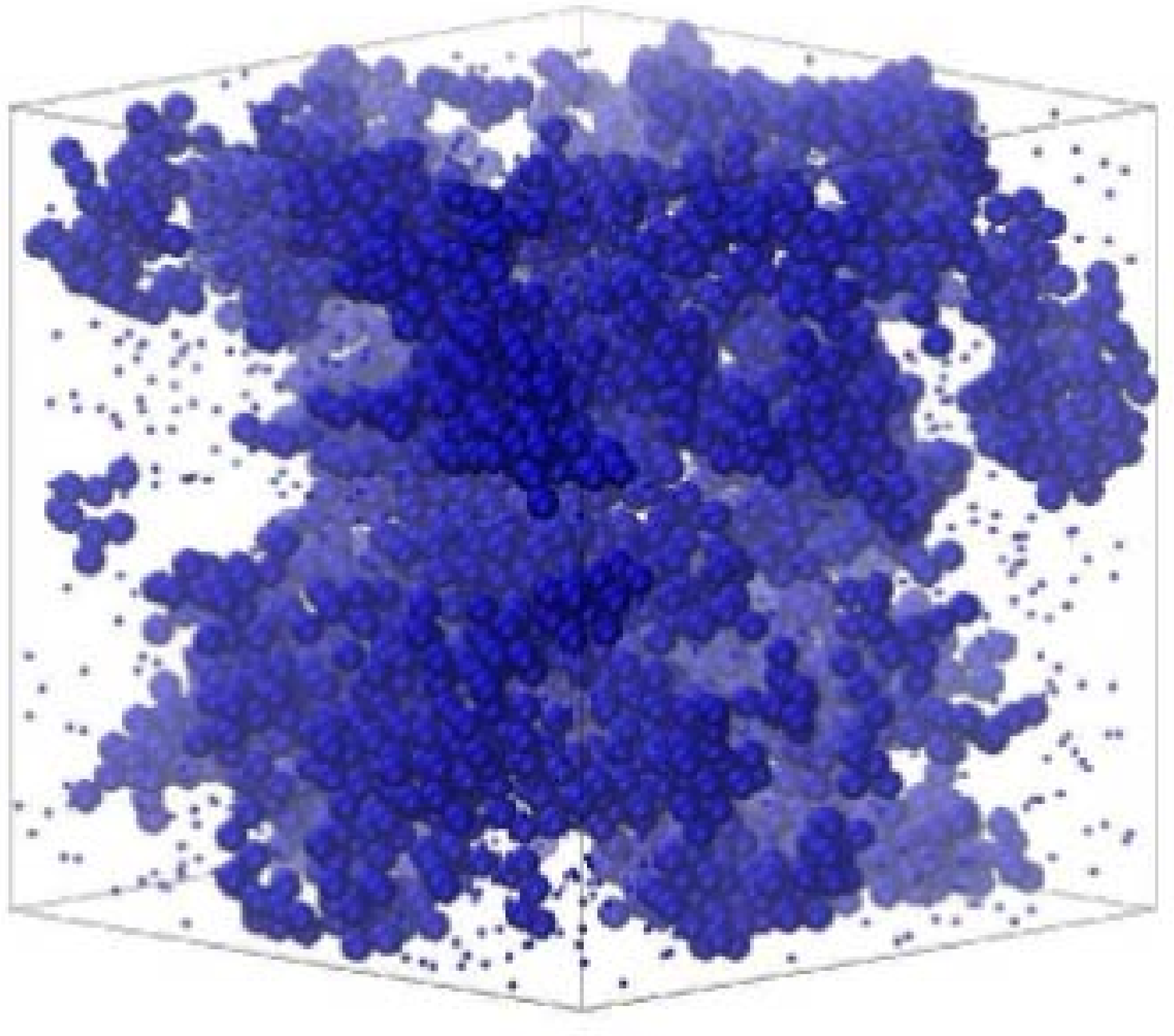}\includegraphics[scale=.29]{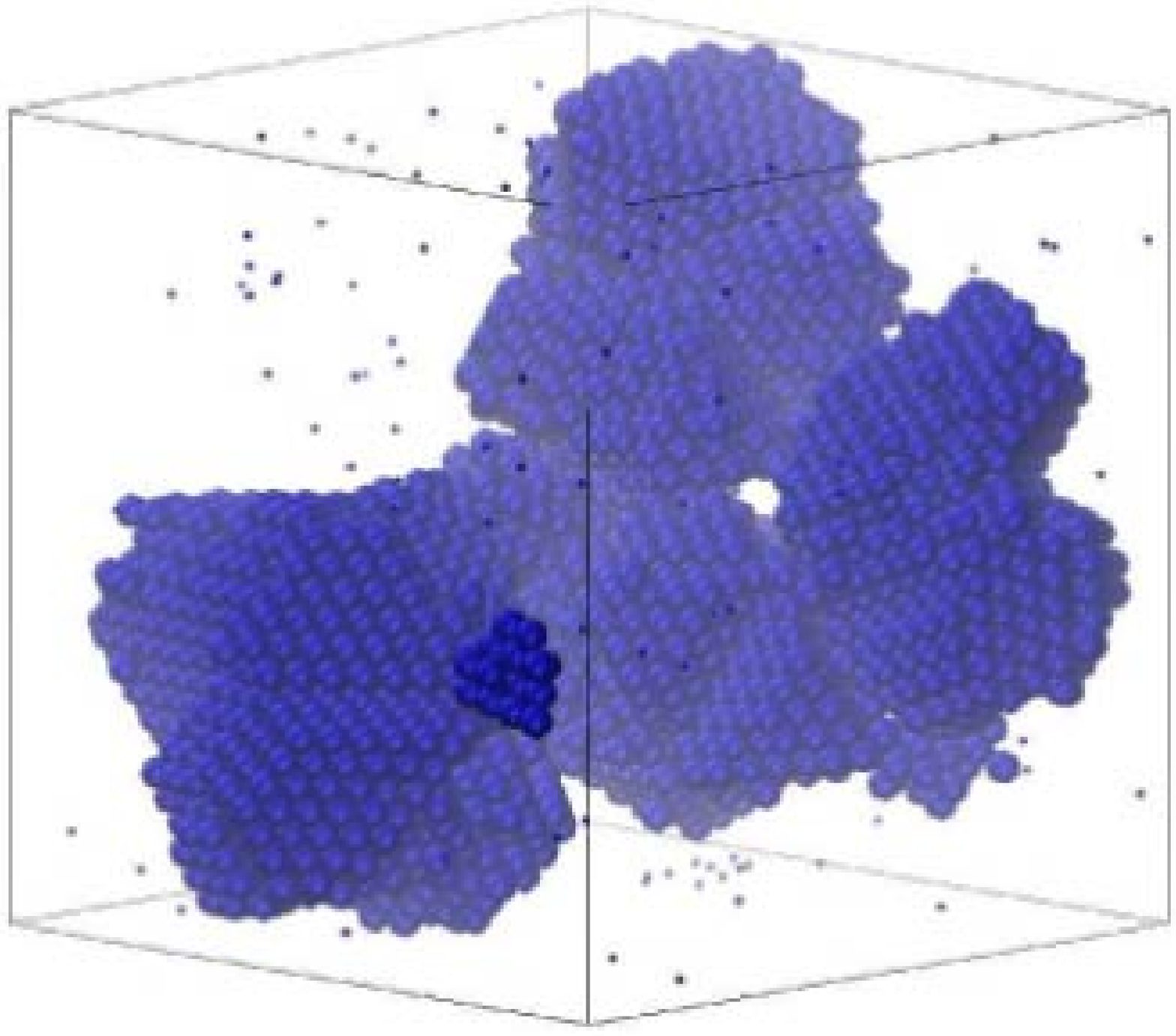}}\\
\hspace{1in}\textbf{(a)\hfill{(b)\hfill(c)}}\hspace{1in}
\caption[Progression of phase separation from homogeneous liquid to
gas-crystal phase coexistence via spinodal
decomposition]{(Color online) Progression of phase separation at $T=0.275$ (point C
in Fig.~\ref{fig:ljph}) from homogeneous liquid to eventual
gas-crystal phase coexistence via the spinodal decomposition due to
crossing the metastable gas-liquid binodal. The system is at
$\phi=15\%$ and is here shown at $t=0$, $300$, and $2000$. For
clarity monomers are represented by smaller spheres.}
\label{fig:ljdyn}
\end{figure*}

The precise location of various phase boundaries has
important implications for the dynamics that might be observed upon
quenching a homogeneous system into different temperature and
density regions of the phase diagram. Interesting out-of-equilibrium
phenomena have been observed experimentally depending on the depth
of the quench, the initial density, and the quench rate. Using
precise characterization of the phase diagram for the LJ ($n=50$)
system of Fig.~\ref{fig:ljph}, we use direct molecular dynamics
simulation to investigate some aspect of this kinetic behavior. For
clarity, we shall refer to the various labeled points in
Fig.~\ref{fig:ljph}. Points A-E are located at a volume fraction
$\phi=\pi\rho/6=15\%$ and at $T=0.50$, $0.35$, $0.275$, $0.20$, and
$0.10$ from A to E, respectively. Points B' and C' refer to volume
fraction $\phi=6.9\%$ and $\phi=4.7\%$ and temperature $T=0.3$ and
$T=0.26$, respectively. The behavior of the system at point A, as
expected, is that of a homogeneous gas as verified by direct
dynamical simulation. No further mention of the behavior in this
region of the phase diagram will be made.

\subsection{Gas-crystal phase coexistence}
Cluster phases in colloidal systems may exist for a variety of
reasons~\cite{sear:2003,malescio:2003,lopes:2002,mladek:2006}.
Perhaps the most ubiquitous reason is the presence of both
short-ranged attractions and long-ranged repulsion. This
repulsion is most commonly thought to arise from excess charge
residing on the colloidal particles. Recent experiments,
however, have suggested that even when charge repulsion is
highly screened, compact clusters may appear and evolve in a
rather stable fashion. Lu \emph{et al}. have recently studied,
via confocal microscopy, the evolution of attractive colloidal
suspensions in a high salt environment~\cite{lu:2006}. They
find evidence for stable, compact clusters of more than 500
particles. This behavior is most prominent at rather weak
attraction strength ($U\sim 2k_BT$) and an intermediate range
of attraction ($\sim0.15\sigma$). Sedgwick \emph{et al}., in
screened lysozyme solution, find evidence for a compact
``bead'' phase in a different region of the phase
diagram~\cite{sedgwick:2005}. The bead phase of Sedgwick
\emph{et al}. appears to exist only close to the metastable
gas-liquid binodal.  Therefore, the ability to locate precise
boundaries in the equilibrium phase diagram is a fundamental
prerequisite for the study of such kinetic behavior. Below, we
undertake a preliminary study of nonequilibrium behavior in the
regions of the phase diagram where such cluster phases have
been found.

In the study of Lu \emph{et al}., the attraction strength where
a cluster phase is observed would appear to be in the broad
gas-crystal coexistence region~\cite{lu:2006}. Further support
for this may be found in the degree of crystallinity seen in
the clusters imaged by Lu \emph{et al}.~\footnote{In the
experiments of Lu \emph{et al.}, surface-induced
crystallization is seen in this regime when clusters contact
the coverslip boundary. P. J. Lu (private communication).}. In
our simulations, we find that, by inserting a small
face-centered-cubic nucleus, the subsequent crystal growth is
rather facile and reaches equilibrium with the gas phase
rapidly. The behavior occurs essentially throughout the
gas-crystal coexistence region above the metastable gas-liquid
binodal. The idea that nucleation and growth of clusters can be
self-limiting, leading to a metastable cluster fluid phase, was
theoretically put forth by Kroy, Cates, and
Poon~\cite{kroy:2004}. Lu \emph{et al}. propose that the
rearrangement of particles at the surface and on the interior
of a cluster may occur on a different time scale than the time
scale on which clusters diffuse away from each other, rendering
clusters long-lived. This notion is similar to that of
Ref.~\cite{kroy:2004}.

To test this kind of hypothesis we study larger system than
that possible by GEMC. We artificially nucleate the solid by
carving out irregular portions of the equilibrium crystal
($\sim 2000$ particles each) allowing the surface of these
clusters to come to equilibrium with the vapor under the
appropriate thermodynamic conditions, and then allowing the
clusters evolve via molecular dynamics.  We have implemented
this procedure at points B and B' of the phase diagram in
Fig.~\ref{fig:ljph}. While this procedure is a crude mimic of
the kinetics of gas-crystal nucleation, it does provide a
useful test of self-limited cluster growth as a stabilizing
mechanism of the cluster phase observed by Lu \emph{et
al.}~\cite{lu:2006}. We find that smaller compact clusters
merge without much impediment upon collision with each other.
This process is depicted in Fig.~\ref{fig:ljclus}. No
trajectories showed cluster dissociation. It should be noted
that the same behavior is observed at higher temperatures as
well, where the initial cluster surface morphologies are
rougher due to a higher equilibrium vapor pressure. We did not
find any evidence of a strong cluster size dependence for this
process. Thus, while we did not make a systematic study of the
effects of quench rate, system size, or polydispersity, one
possibility for the cluster fluid phase observed by Lu \emph{et
al}. is simply a slowly evolving solid nucleation and growth
process~\footnotemark[\value{footnote}]. Interestingly,
Sedgwick \emph{et al}. find that the crystal phase is observed
for a wide range of volume fractions and temperatures
effectively in the gap between gas-solid and metastable
gas-liquid coexistence. This is consistent with our findings in
the analogous region of the phase diagram.

A final possibility that should be mentioned with regard to the
cluster phase observed by Lu \emph{et al}. is accumulation
repulsion~\cite{louis:2002,likos:2005}. Even in the absence of
charge, it is possible that the neutral added polymer may
induce an effective repulsion between colloidal spheres, due to
its enhanced concentration at the surface of the colloids. It
would appear unlikely that this repulsion is sufficient to
render clusters thermodynamically stable in the system studied
by Lu \emph{et al}. However, it is possible that this
accumulation repulsion could significantly enhance metastable
cluster lifetimes.  To the best of our knowledge, a systematic
study of nucleation and growth kinetics as a function of
repulsion strength and range has never been performed (see,
however,~\cite{pini:2000}). Such a study would be quite useful
in understanding the role of the added polymer on the stability
of the cluster phase observed by Lu \emph{et al}.

\subsection{Metastable gas-liquid phase separation}

\begin{figure}
\center{\includegraphics[width=0.8\columnwidth]{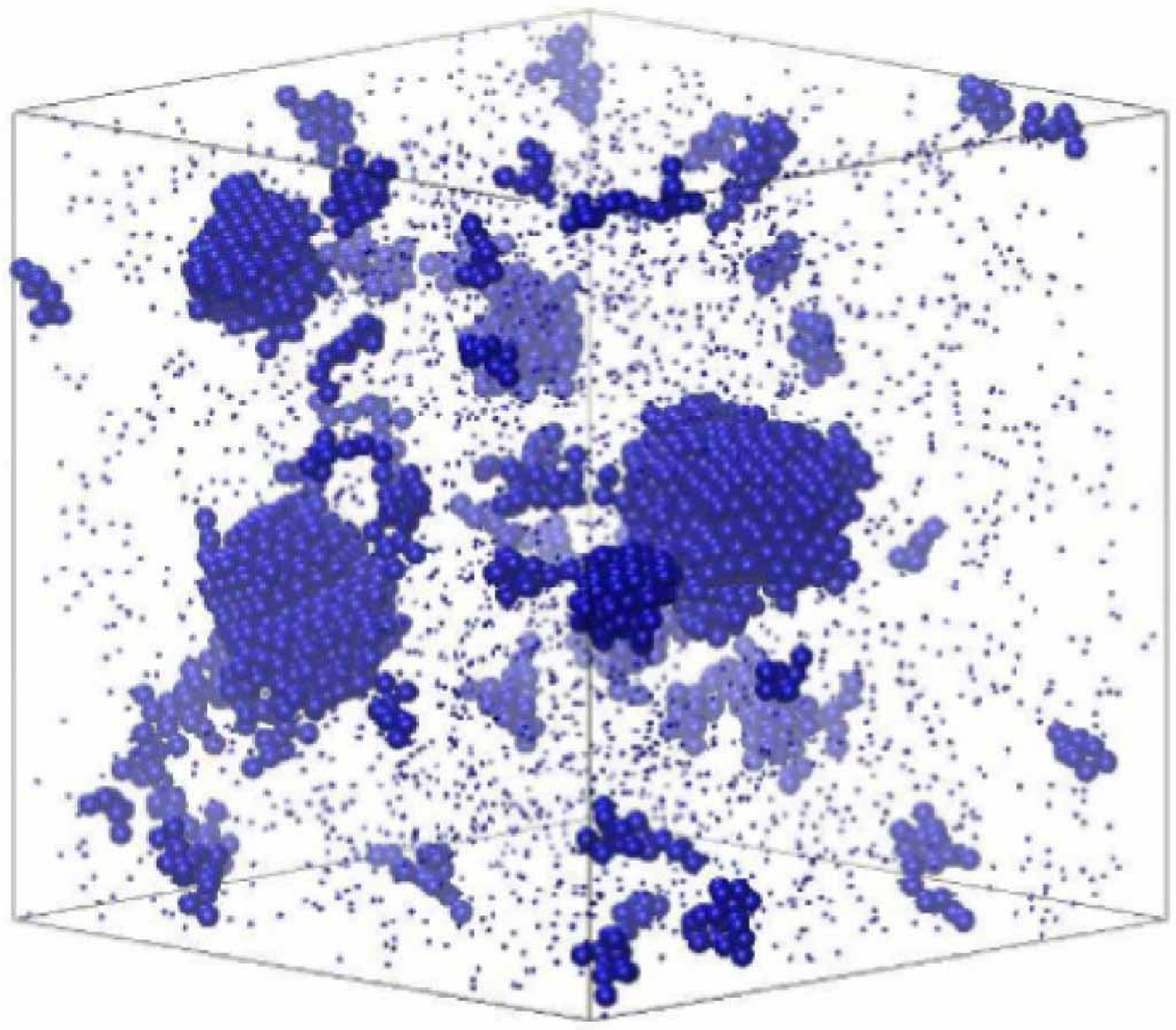}\\\includegraphics[scale=.3]{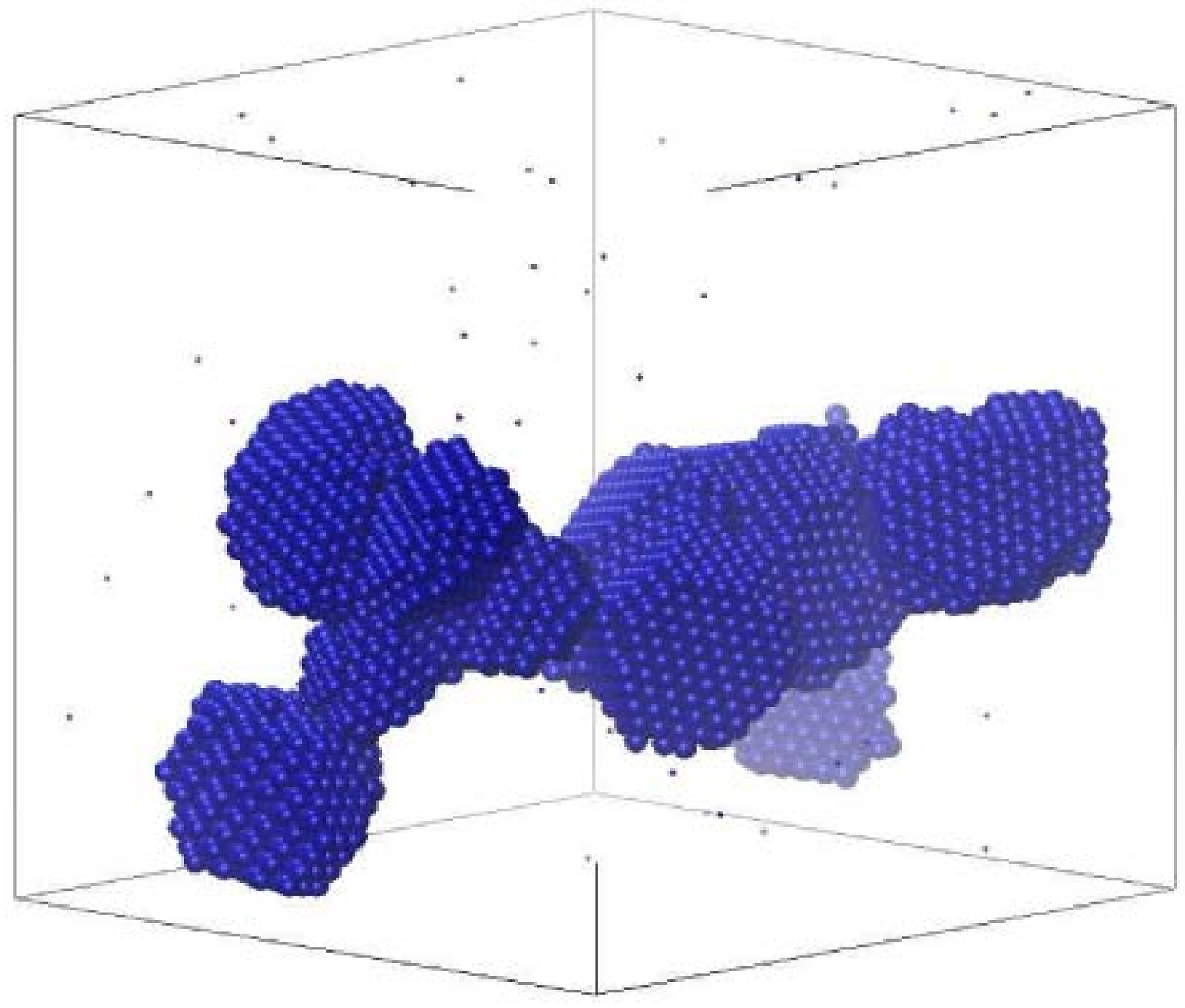}\\
} \caption[Evolution of nucleated metastable liquid droplets]{(Color online) Fate
of nucleated liquid droplets upon a rapid quench to point C' of
Fig.~\ref{fig:ljph}. Upper panel shows early stages
($t=2.2\times10^3$) and lower panel shows later stages of
solidification process ($t=3.9\times10^4$). For clarity particles
that are part of small clusters ($n<6$) and monomers are represented
with smaller spheres.} \label{fig:ljbeads}
\end{figure}

\begin{figure*}
\center{\includegraphics[scale=.3]{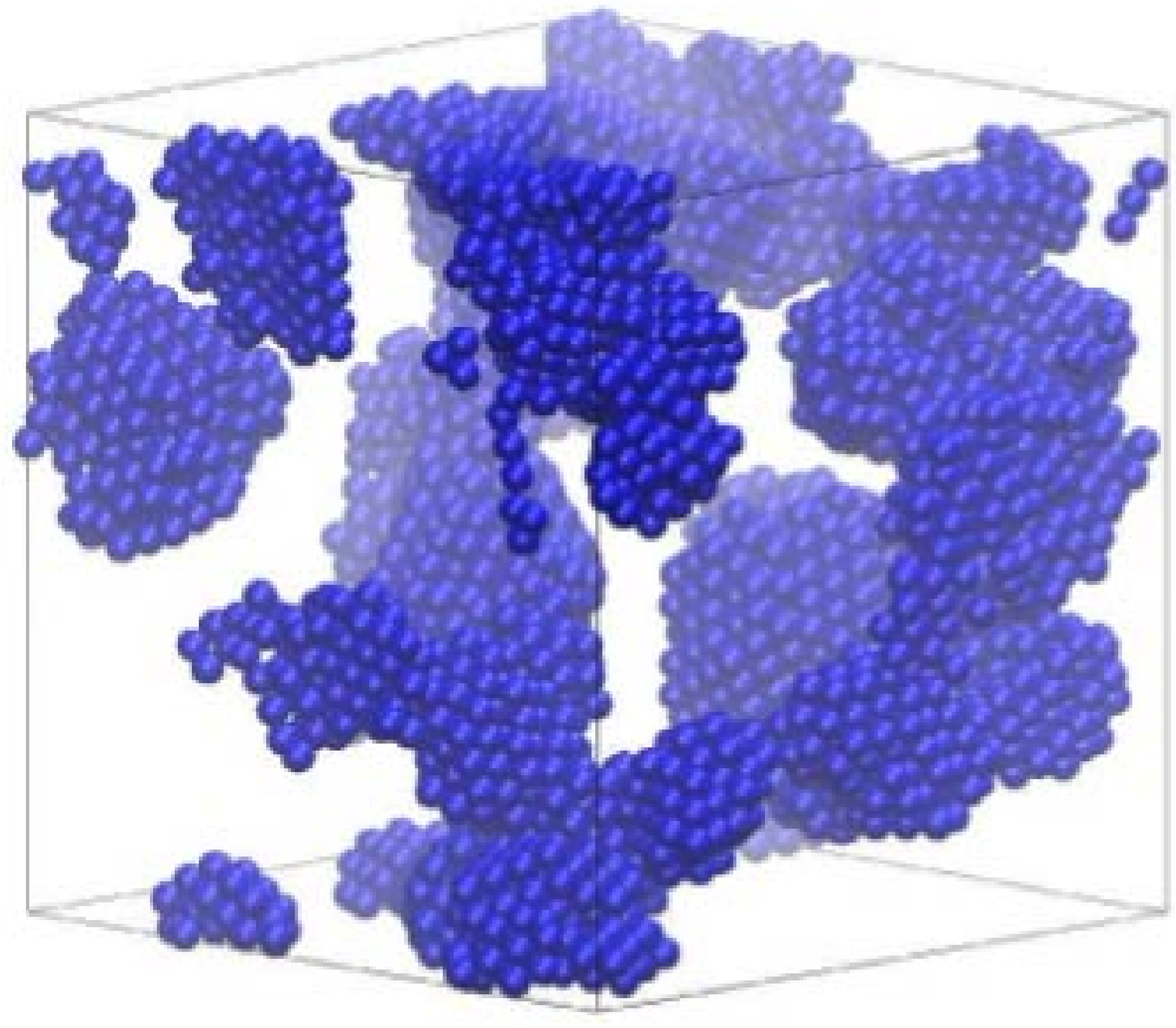}\includegraphics[scale=.305]{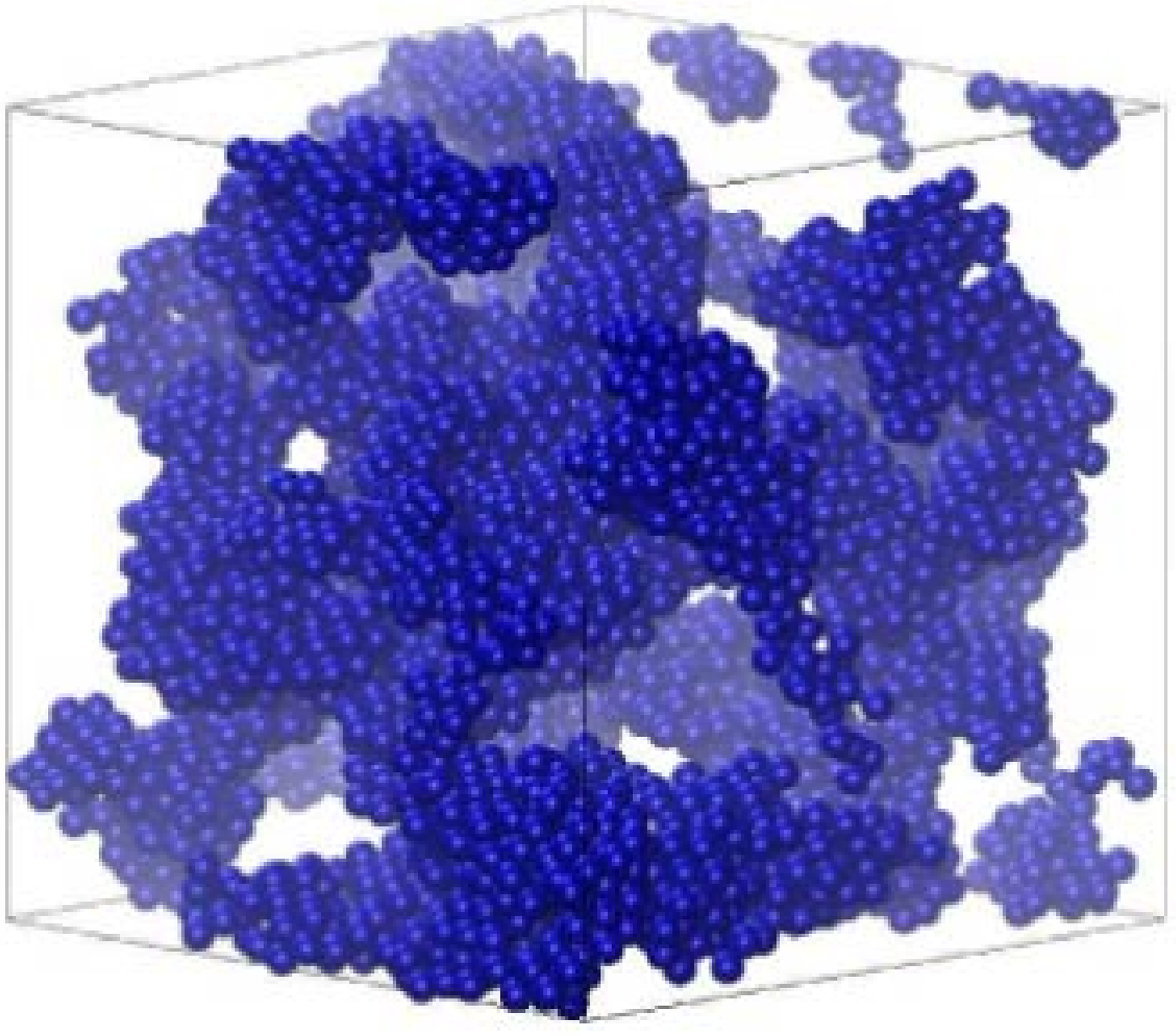}\includegraphics[scale=.3]{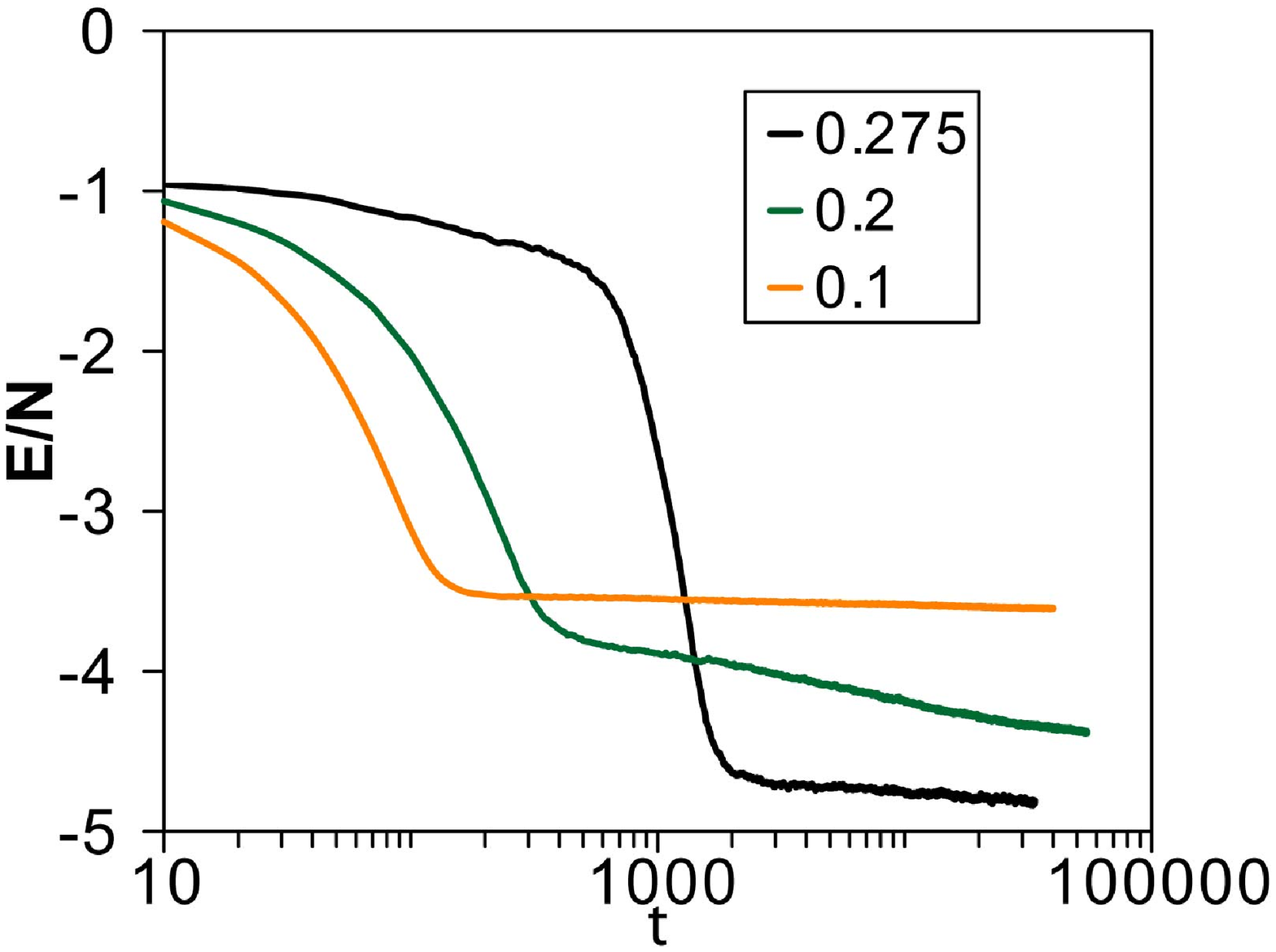}}
\vspace{-0.2in}\\
\hspace{1in}\textbf{(a)\hfill{(b)\hfill(c)}\hspace{1in}}
\caption[Gel structures and potential energy evolution at
different quench depths]{(Color online) Comparison of
solidified structures of LJ systems with $n=50$ at $\phi=15\%$
quenched under the spinodal line (\textbf{a}) to $T=0.2$ at
$t=7.3\times 10^3$ and (\textbf{b}) to $T=0.10$ at $t=7.3\times
10^3$, which correspond, respectively, to points D and E in
Fig.~\ref{fig:ljph}. (\textbf{c}) Evolution of the energy per
particle for configurations quenched at points C, D, and E.
Note that the configuration at point C fully crystallizes.}
\label{fig:ljperc}
\end{figure*}

A generic feature of systems with short-ranged attractions is
the existence of a metastable gas-liquid binodal buried below
the gas-crystal coexistence line.  This feature gives rise to
another possible phase separation mechanism. In particular, a
two-step nucleation process, where the first step of phase
separation is the formation of a higher density liquid and the
subsequent step involves the transformation of the high density
liquid into a crystal, arises. This two-step nucleation process
can proceed by a lower free energy pathway than that of
classical nucleation, thus accelerating the rate of crystal
formation. At low densities, just below the metastable
gas-liquid binodal, interesting nonequilibrium behavior has
been reported.

In particular, it is in the vicinity of the metastable
gas-liquid coexistence line that the bead phase of Sedgwick
\emph{et al}. is observed~\cite{sedgwick:2005}. Thus, in this
section, we investigate the dynamics of domain growth at two
volume fractions (point C and C' in Fig.~\ref{fig:ljph}) just
below the metastable gas-liquid binodal.

When our system is quenched to point C of Fig.~\ref{fig:ljph},
rapid nucleation of ramified liquid regions, followed by
solidification occurs. Figure~\ref{fig:ljdyn} shows this
two-step process which is clear upon visual inspection, and may
also be observed in the time evolution of the energy of the
system per particle, as shown in Fig.~\ref{fig:ljperc}. For
lower density quenches (point C'), liquid beads nucleate
quickly. The evolution of the low density droplet phase then
occurs via cluster diffusion. Generically, the droplets begin
to crystalize before they coalesce, as seen in
Fig.~\ref{fig:ljbeads}. The crystal growth process thus takes
place by cluster coalescence, and it strikingly similar to the
growth of the crystal above the metastable gas-liquid binodal.
It should be noted that the entire process illustrated in
Fig.~\ref{fig:ljbeads} occurs spontaneously, in contrast to the
\emph{ad hoc} seeding procedure that we have performed to
illustrate gas-crystal phase separation at higher temperatures
above the gas-liquid binodal. Indeed, this is possible due to
the rapid formation of liquid beads in this region of the phase
diagram.

Clearly, we do not observe that liquid beads are long-lived
along the low temperature side of the metastable gas-crystal
binodal. This appears to contrast with the results of Sedgwick
\emph{et al.}~\cite{sedgwick:2005}. It is interesting to note,
however, that recent confocal microscopy studies of Lu \emph{et
al}.~\footnote{P. J. Lu (private communication).} at
$\phi=0.05$ do show clear coalescence of clusters. There are
several reasons why the final stages of nucleation might occur
more slowly in the experimental system of Sedgwick \emph{et
al}. than in our simulations. First, our system is
monodisperse, which facilitates the crystallization process.
Crystal beads may coalesce more easily due to surface faceting,
providing a more regular contact area between clusters. Second,
the lysozyme units of Sedgwick \emph{et al}. carry some
residual charge which may hinder crystal formation. Third,
sedimentation may play some role, and is clearly not modeled in
our system. It should also be mentioned that the quench rate
may play a significant role in the formation of such phases, as
Sedgwick \emph{et al}. have mentioned. Although we have not
made a systematic study of the effects of quench rate, our
preliminary studies show the same behavior for systems rapidly
quenched to point C' and those that reach metastable
equilibrium just above point C' (homogeneous fluid) and are
then slowly quenched to point C'. Thus, we find no clear
evidence of quench rate dependence in our system in this region
of the phase diagram.

\subsection{Deep quenches: gels}

\begin{figure}
\center{\includegraphics[width=0.9\columnwidth]{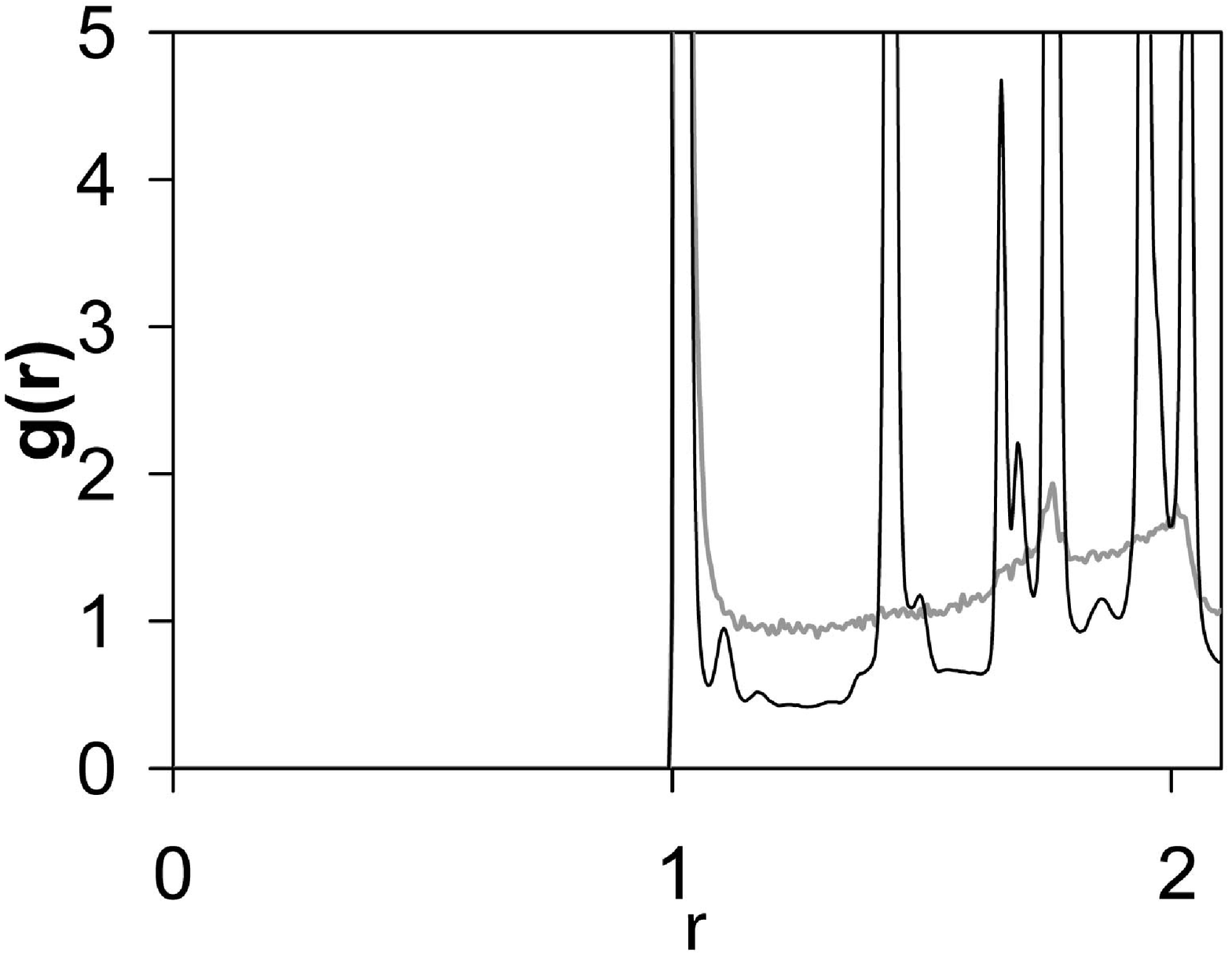}\\
\includegraphics[width=0.9\columnwidth]{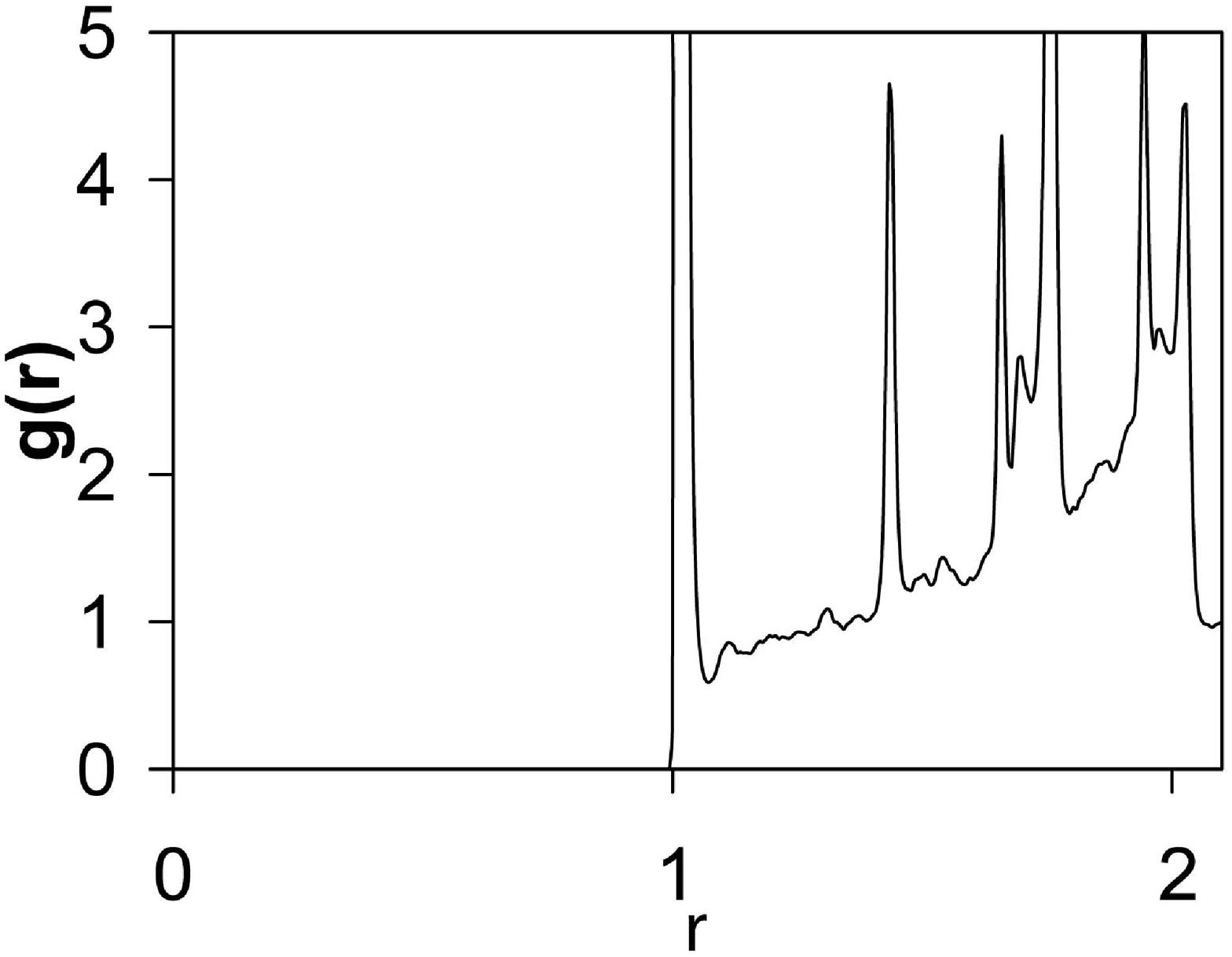}}
\caption[Radial distribution function for quenched
systems]{Radial distribution function for system quenched (a)
to point D ($t\sim0$ in gray and at $t\sim5\times 10^4$ in
black) and (b) to point E ($t\sim1\times 10^3$).}
\label{fig:ljstruc}
\end{figure}

When systems with short-ranged attractions are quenched below
the metastable gas-liquid binodal curve, phase separation
characterized by large-scale fluctuations sets in.  At short
times, the system segregates into liquid-rich and liquid-poor
regions. Since the gas-liquid critical point is buried below
the gas-solid coexistence line, the fluid phase is metastable
with respect to the crystal. Thus, an additional process, the
nucleation of the stable crystal phase from the metastable
fluid competes with spinodal decomposition, and the time
dependence of the nonequilibrium process is governed by a
subtle interplay of factors, such as the relative rates of
solid nucleation within the liquid to the rate of the initial
decomposition process~\cite{tenwolde:1997}. If either
crystallization or vitrification occurs locally inside the
dense component before the phase separation process is
complete, the long wavelength segregation of phases slows down
considerably.  On time scales relevant for colloidal
experiments, porous solids that bear the
imprint of this arrested phase separation appear~\cite{cates:2004,%
verhaegh:1997,manley:2005,foffi:2005,foffi:2005b,lodge:1999}. Since many recent
experimental and computational studies of this route to colloidal
gelation have been carried out, we confine ourselves here to some
rather qualitative features of this process in this section.  The
distinction between the past work and the work presented here is
that a rather precise characterization of such behavior with respect
to the location of various phase boundaries (here the location
of the metastable gas-liquid binodal) may be made.

In Fig.~\ref{fig:ljperc}(a) and Fig.~\ref{fig:ljperc}(b) two
snapshots of the intermediate evolution of the phase separation
process are shown.  These snapshots correspond to direct
quenches to points D and E in Fig.~\ref{fig:ljph},
respectively. While quenches just below the metastable
gas-liquid binodal allow for facile crystallization, the
evolution of the system for deeper quenches is anomalously
slow.  This is clearly illustrated by the time dependence of
the energy per particle as illustrated in
Fig.~\ref{fig:ljperc}(c). For the intermediate quench (D) there
is residual evolution of the energy, while for deeper quenches
(E) near complete arrest is observed.

Given the weaker bonding relative to temperature at point D, it
is expected that the system may explore deeper metastable
states during its slow evolutions.  This is clearly seen in
Fig.~\ref{fig:ljperc}(c). Correspondingly, a greater degree of
crystallinity is expected in the interior of the porous solid
when compared to deeper quenched structures that arrest at
earlier times during their structural evolution.  Indeed, as
shown in Fig.~\ref{fig:ljstruc} more resolved crystal peaks are
observed in the radial distribution function $g(r)$ for more
shallow quenches.

A topic of great interest of late is the arrest of phase
separation by vitrification as opposed to crystallization.  In
the investigation here, the monodisperse nature of the sample
favors crystallization. By making the sample polydisperse,
arrested phase separation will occur via a local glass
transition of the dense phase~\cite{foffi:2005,foffi:2005b}.
The characterization of the precise location of such a glass
transition is greatly complicated by a variety of features,
including the explicitly nonequilibrium nature of the process
and the fact that the effective density of the component
undergoing the glass transition is the thermodynamically
defined bulk density. The use of GEMC enabling essentially
exact determination of stable and metastable phase behavior,
perhaps combined with the ideas of Ref.~\cite{sastry:2000}
might lead to a more precise characterization of the notion and
location of a ``glass transition'' under the metastable
gas-liquid binodal residing in the expanded parameter space
that includes quench dependent and thermodynamic parameters.

\section{Conclusions}
\label{sect:conclu} In this work we have demonstrated the
general utility and feasibility of GEMC for the study of the
phase behavior of systems with rapidly varying, short-ranged
attractions.  Thus, rather exhaustive study of both the
equilibrium and nonequilibrium behavior of such systems is
possible for generic potentials and with precise reference for
various phase boundaries.  Here, we have mainly made comparison
to the predictions of the SCOZA approach and made a preliminary
investigation of dynamics.  In light of recent experiments that
show interesting nonequilibrium phases in systems with
predominantly short-ranged attractions, more work should be
performed to understand the role of quench rate,
polydispersity, gravity, hydrodynamics, and other factors that
may influence the dynamical behavior of such systems. Another
avenue worthy of study are systems that possess long-ranged
repulsion in addition to short-ranged attraction.  Such systems
have been the focus of intense recent study.  Due to the
competing lengthscales in these systems, implementation of the
GEMC method is rather demanding for larger systems. A
preliminary GEMC investigation of gelation and microphase
separation in such systems will be presented in a future
publication~\cite{charbonneau:2006}.

\begin{acknowledgments}
This work was supported in part by the NSF-0134969 and
FQRNT-91389 (to P.C.) grants. We would like to thank P. Lu and
K. Miyazaki for helpful discussions and comments.
\end{acknowledgments}

\bibliography{AttGel}
\end{document}